%% file: main.tex
\documentclass[12pt]{article}
\usepackage{amsmath}
\usepackage{graphicx}
\usepackage{enumerate}
\usepackage{url} 

\usepackage{qz_package}
\DeclareRobustCommand{\qzz}[2]{#2}
\soulregister{\qzz}{2}

\newcommand{\blind}{1}

\begin{document}

\def\spacingset#1{\renewcommand{\baselinestretch}%
{#1}\small\normalsize} \spacingset{1}


\if1\blind
{
  \title{\bf What is a randomization test?}
  \author{Yao Zhang \hspace{.2cm}\\
    Department of Statistics,\\
    Stanford University\\
    and \\
    Qingyuan Zhao\thanks{QZ was partly supported by a research grant from
      the Isaac Newton Trust and EP/V049968/1 from the Engineering and
      Physical Sciences Research Council (EPSRC).} \\
    Statistical Laboratory, University of Cambridge}
  \maketitle
} \fi

\if0\blind
{
  \bigskip
  \bigskip
  \bigskip
  \begin{center}
    {\LARGE\bf What is a randomization test?}
\end{center}
  \medskip
} \fi

\bigskip
\begin{abstract}
The meaning of randomization tests has become obscure in statistics
  education and practice over the last century. This article makes a
  fresh attempt at rectifying this core concept of statistics.
 A new term---``quasi-randomization test''---is introduced to define
 significance tests based on theoretical models and distinguish these
 tests from the ``randomization tests'' based on the physical act of
 randomization. The practical importance of this distinction is
 illustrated through a real stepped-wedge cluster-randomized
 trial. Building on  the recent literature on randomization inference,
 a general framework of conditional randomization tests is developed
 and some practical methods to construct conditioning events are
 given. The proposed terminology and framework are then applied to
 understand several widely used (quasi-)randomization tests, including
 Fisher's exact test, permutation tests for
 treatment effect, quasi-randomization tests for independence and
 conditional independence, adaptive randomization, and conformal
 prediction.
\end{abstract}

\noindent%
\vfill

\newpage
\spacingset{1.9} 

\input{1_Introduction_revised}

\section{An illustrative example: The Australia weekend health
  services disinvestment trial}
\label{sec:example}

\input{2_Australian_Trial}

\section{A general theory for randomization tests}
\label{sec:single-crt}

\input{3_Single_CRT_revised}

\section{Practical methods of CRTs}
\label{sec:practical-methods}

\input{4_Practical_Methods}

\section{Examples of (quasi-)randomization tests}
\label{sec:examples}

\input{5_Examples_revised}

\section{Discussion}
\label{sec:discussion}

\input{6_Discussion_Revised}

\newpage
\small
\spacingset{1}
\bibliographystyle{Chicago}
\bibliography{./swd.bib}

\newpage
\appendix
\section*{Appendix}

\input{appendix}

\end{document}

%% file: 1_Introduction_revised.tex
\section{Introduction}\label{new_intro}

Randomization is one of the oldest and most important ideas in
statistics, playing several roles in experimental designs and inference
\citep{cox09_random_desig_exper}. Randomization tests
were introduced by \citet[Chapter
21]{fisher1935design}\footnote{\citet{onghena2017randomization}
  argued that
  Fisher did not embrace the randomization model in that Chapter and
  should not be given the credit of discovering the randomization
  test.} to
substitute Student's $t$-tests when normality does not hold, and to restore
randomization as ``the physical basis of the validity of statistical tests''. This idea was immediately extended by
\citet{pitman1937significance}, \citet{welch1937z},
\citet{wilcoxon1945individual},
\citet{kempthorne1952design}, and many others.

An appealing property of randomization tests is that they have exact control of the nominal type I error
rate in finite samples without relying on any distributional assumptions. This
is particularly attractive in modern statistical applications that
involve arbitrarily complex sampling distributions. Recently, there
has been a rejuvenated interest in randomization tests in
several areas of statistics, including testing associations in genomics
\citep{bates2020causal,efron2001empirical}, testing conditional
independence \citep{candes2018panning,berrett2020conditional},
conformal inference for machine learning methods
\citep{lei13distribution,vovk2005algorithmic},
analysis of complex experimental designs
\citep{ji2017randomization,morgan2012rerandomization}, evidence
factors for observational studies
\citep{karmakar2019integrating,rosenbaum2010evidence,rosenbaum2017general},
and causal inference with interference
\citep{athey2018exact,basse2019randomization}.

Along with its popularity, the term ``randomization test'' is
increasingly used in statistics and its applications, but
unfortunately often not to represent what it means originally.
For example, at the time of writing Wikipedia redirects ``randomization
test'' to a page titled ``Resampling (statistics)'' and describes it
alongside bootstrapping, jackknifing, and subsampling. The terms
``randomization test'' and ``permutation test'' are often used
interchangeably, which causes a great deal of confusion. A common
belief is that randomization tests rely on certain kinds of group
structure or exchangeability
\citep{lehmann2006testing,southworth2009properties,rosenbaum2017general}. This
led some authors to categorize randomization tests as a special case
of ``permutation tests'' \qzz{}{\citep{ernst2004permutation} or vice
  versa \citep[p.\ 632]{lehmann2006testing}.}
Furthermore, some authors started to use new and, in our opinion,
redundant terminology. An example is
``rerandomization test''
\citep{brillinger1978role,gabriel83_reran_infer_regres_shift_effec},
which is nothing more than an usual randomization test and confuses with
a technique called ``rerandomization'' that is useful for improving
covariate balance \citep{morgan2012rerandomization}. For a
historical clarification on the terminology, we refer the
readers to \citet{onghena2017randomization}.

\qzz{}{The main objective of this paper is to give a clear-cut
  formulation of the randomization test, so it can be distinguished
  from closely related concepts. Our formulation follows from the work
  of
  \citet{rubin80comment}, \citet{rosenbaum02_covar_adjus_random_exper_obser_studies},
  \citet{basse2019randomization}, and many others in causal inference
  based on the potential outcomes model first conceived by
  \citet{neyman1923application}. In fact, this was also the model
  adopted in the first line of works on the randomization tests by
  \citet{pitman1937significance}, \citet{welch1937z}, and
  \citet{kempthorne1952design}. However, as the fields of survey
  sampling and experimental design grew apart since the
  1930s,\footnote{This may have been caused by a controversy between
    Fisher and Neyman over a 1935 paper by Neyman; see \citet[Section
    4.1]{sabbaghi14_commen_neyman_fisher_contr_its_conseq}.} and
  because randomization tests often take simpler forms under the convenient
  exchangeability assumption, the popularity of the potential outcomes
  model had dwindled until \citet{rubin1974estimating} introduced it
  to observational studies. In consequence, most contemporary
  statisticians are not familiar with this approach that can
  give a more precise characterization of randomization test and
  distinguish it from related ideas. Our discussion below will
  focus on the conceptual differences and key statistical ideas;
  detailed implementations and examples of randomization tests (and
  related tests) can be found in the references in this article and
  the book by \citet{edgington2007randomization}.}

\subsection{Randomization tests vs.\ Permutation tests}
\label{sec:random-vs-permutation}

Many authors have repeatedly pointed out that randomization
tests and permutation tests are based on different assumptions and the
distinguishing these terms is crucial. On one hand, randomization
tests are based on ``experimental randomization''
\citep{kempthorne69_behav_some_signif_tests_under_exper_random,rosenberger18_random},
``random assignment'' \citep{onghena2017randomization}, or the
so-called ``randomization model''
\citep{ernst2004permutation,lehmann1975nonparametrics}. \qzz{One}{On} the other
hand, permutation tests are based on ``random sampling''
\citep{kempthorne69_behav_some_signif_tests_under_exper_random,onghena2017randomization},
the so-called ``population model''
\citep{ernst2004permutation,lehmann1975nonparametrics,rosenberger18_random},
\qzz{}{or certain algebraic group structures
\citep{lehmann2006testing,southworth2009properties,hemerik20_anoth_look_at_lady_tastin}}. Moreover,
unlike permutation tests, the validity of randomization tests is not
based on the exchangeability assumption
\citep{kempthorne69_behav_some_signif_tests_under_exper_random,onghena2017randomization,rosenberger18_random,hemerik20_anoth_look_at_lady_tastin}. Despite
these suggestions, randomization test and permutation test are
seldomly distinguished in practice.

The fundamental reason behind this confusing nomenclature is that a
randomization test coincides with a permutation test in the simplest example where
half of the experimental units are randomized to treatment and half to
control. The coincidence is caused by the fact that all
permutations of the treatment assignment are equally likely to
realize in the assignment distribution. As this is usually the first example in a lecture or article
where randomization tests or permutation tests are introduced, it is
understandable that many think the basic ideas behind the two tests
are the same.

The reconciliation, we believe, lies precisely in the
names of these tests. Randomization refers to a physical action that makes the
treatment assignment random, while permutation refers to a step of an
algorithm that computes the significance level of a test.
The former emphasizes the basis of inference, while the latter
emphasizes the algorithmic side of inference, so neither
``randomization test'' or ``permutation test'' subsumes the other. In
the example above, we can use either term to refer to the same test, but the
name ``randomization test'' is preferable as it provides more
information about the context of the problem.

\subsection{Randomization tests vs.\ Quasi-randomization tests}

To further clarify the distinction between randomization tests and
permuation tests, we believe it is helpful to introduce a new
term---``quasi-randomization tests''. It refers to any test that is not
based on the physical act of randomization. It is exactly the
complement of randomization tests. A test can then be characterized \qzz{by
two aspects}{in two dimensions}:
(1) whether it is based on the physical act of randomization;
and (2) how it is computed (using permutations, resampling, or
  distributional models).
With this in mind, we can now distinguish two tests that are computationally
identical (in the sense that the same acceptance/rejection
decision is always reached given the same observed data) based on their
underlying assumptions.


To illustrate this, consider permutation test in the following two
scenarios. In the first scenario, an even number of units
are paired before being randomized to receive one of two treatments, with
exactly one unit in each pair receiving each treatment. In the
second scenario, the units are observed (but not randomized)
and we pair each unit receiving the first treatment with a different
unit receiving the second treatment. To test the null hypothesis that
the two treatments have no difference on any unit, we can permute treatment
assignments within the pairs. Although these tests are
computationally identical, the same conclusion (e.g.\ rejecting the null
hypothesis) from them may carry very different weight. The second test
relies on the assumption that the two units in the same pair are
indeed exchangeable besides their treatment
status. This assumption can be easily violated if the units are
different in some way; even if they look comparable in every
way we can think of now, someone in the future may discover an
overlooked distinction. Randomization plays a crucial role in hedging
against such a possibility
\citep{kempthorne69_behav_some_signif_tests_under_exper_random,marks03_rigor_uncer}. In
the new terminology we propose, both tests are permutation tests, but the
first is a randomization test and the second is a quasi-randomization
test. \qzz{}{That is, even if a randomization test and a
  quasi-randomization test are algorithmically identical, they have
  entirely different inferential basis and thus must be
  distinguished.}

The distinction between a randomization test and a quasi-randomization
test is intimately related to causal inference using experimental and
observational data. Our nomenclature is motivated by term
``quasi-experiment'' coined by \citet{campbell1963experimental} to
refer to an observational
study that is designed to estimate the causal impact of an
intervention. Since then, this term has been widely used in social
science \citep{cook2002experimental}.

\subsection{Randomness used in a randomization test}

At this point, our answer to the question in the title of this article
should already be clear. A randomization test is precisely what
its name suggests---a hypothesis test\footnote{In this article we
  use the terms ``significance test'' and ``hypothesis test''
  interchangeably. Some authors argued that we should also distinguish
  a significane test, ``as a conclusion or condensation device'', from
  a hypothesis test ``as a decision device''
  \citep{kempthorne69_behav_some_signif_tests_under_exper_random}.
This is closely related to the ``inductive inference'' vs.\
  ``inductive behaviour'' debate between Fisher and Neyman
  \citep{lehmann93_fisher_neyman_pears_theor_testin_hypot}.}
based on randomization and
nothing more than randomization. But what does ``based on
randomization'' exactly mean?
To answer this question, it is helpful to consider counterfactual
versions of the data. In the causal inference literature, this is
known as the Neyman-Rubin model \citep{holland1986statistics}, which
postulates the existence of a potential outcome (or counterfactual)
for every possible realization
of the treatment assignment. Broadly speaking, a ``treatment
assignment'' can be anything that is randomized in an experiment (so
not an actual treatment), while the ``outcome'' includes everything
observed after randomization.
To clarify the nature of randomization tests, we separate the
randomness in data and statistical tests into
\begin{enumerate}
\item Randomness introduced by the nature in the
  potential outcomes;
\item Randomness introduced by the experimenter (e.g., drawing balls
  from an urn); 
\item Randomness introduced by the analyst, which is optional.
\end{enumerate}
Using this trichotomy, a randomization test can be understood as a
hypothesis test that conditions on the potential
outcomes and obtains the sampling distribution (often called the
\emph{randomization distribution}) using the
second and third sources of randomness. A
randomization test is based solely on the randomness introduced by
humans (experimenters and/or analysts), thereby providing a coherent
logic of scientific induction \citep{fisher1956statistical}.

We would like to make two comments on the definitions above. First,
the notion of potential outcomes, first introduced by Neyman in the
context of randomized agricultural experiments, was not uncommon in
the description of randomization tests. This was often implicit, but
the seminal paper by \citet{welch1937z} used potential
outcomes to clarify that randomization test is applicable to Fisher's sharp
null  rather than Neyman's null concerning the
average treatment effect. Second,
the difference between randomization before and after an
experiment is also well recognized
\citep{basu1980randomization,kempthorne69_behav_some_signif_tests_under_exper_random}.

Much of the recent literature on randomization tests is motivated by
the interference (cross-unit effect) problem in causal
inference. A key feature of the interference problem is that the null
hypothesis is only ``partially sharp'', another term we coin in this
article to refer to the phenomenon that the potential outcomes are not
always imputable under all possible treatment assignments. In
consequence, a randomization test that uses all the randomness
introduced by the experimenter may be uncomputable. A general solution
to this problem is conditioning on some carefully constructed
events of the treatment assignment.


\subsection{An overview of the article}
\label{sec:outline-article}

\Cref{sec:example} investigates a real cluster-randomized controlled
trial using (quasi-)randomization tests that are based
on different assumptions about the data. \Cref{sec:single-crt}
develops an overarching theory for (conditional) randomization
tests by generalizing the classical Neyman-Rubin causal model.
The usage of the potential outcome notation allows us to give a
precise definition of randomization test. \Cref{sec:practical-methods}
then reviews some practical methods
to construct conditional randomization tests. \Cref{sec:examples}
discusses some quasi-randomization tests in the recent literature,
including tests for (conditional) independence and conformal
prediction. Finally, \Cref{sec:discussion} concludes the paper with
some further discussion.

\qzz{}{
\textbf{Notation.} We use
calligraphic letters for sets,
boldface letters for vectors, upper-case letters for
random quantities, and lower-case letters for fixed quantities.
We use a single
integer in a pair of square brackets as a shorthand notation for the
indexing set from $1$: $[N] = \{1,\dotsc,N\}$. We use set-valued
subscript to denote a sub-vector; for example, $\bm Y_{\{2,4,6\}} =
(Y_2,Y_4,Y_6)$. We use $X \independent Y$ to represent the random
variable $X$ is independent of the random variable $Y$.
}


%% file: 2_Australian_Trial.tex
\qzz{}{We first illustrate the conceptual and practical differences of
randomization and quasi-randomization tests through a real data example.}

\subsection{\qzz{}{Trial background}}

\citet{haines17_impac_disin_from_weeken_allied} reported
the results of a cluster randomized controlled trial about the impact
of disinvestment from weekend allied health services across acute
medical and surgical wards. The trial consisted of two phases---in the
first phase, the original weekend allied health service model
was terminated, and in the second phase, a newly developed model was
instated. The trial involved 12 hospital wards in 2 hospitals in
Melbourne, Australia. As our main purpose is to demonstrate the
distinction between randomization and
quasi-randomisation tests, we will focus on the first phase of the
trial and 6 wards in the Dandenong Hospital.
The original article investigated a number of
patient outcomes; below we will just focus on patient length of stay
after a log transformation.

\begin{figure}[h]
  \centering
  \includegraphics[width = 0.75\textwidth]{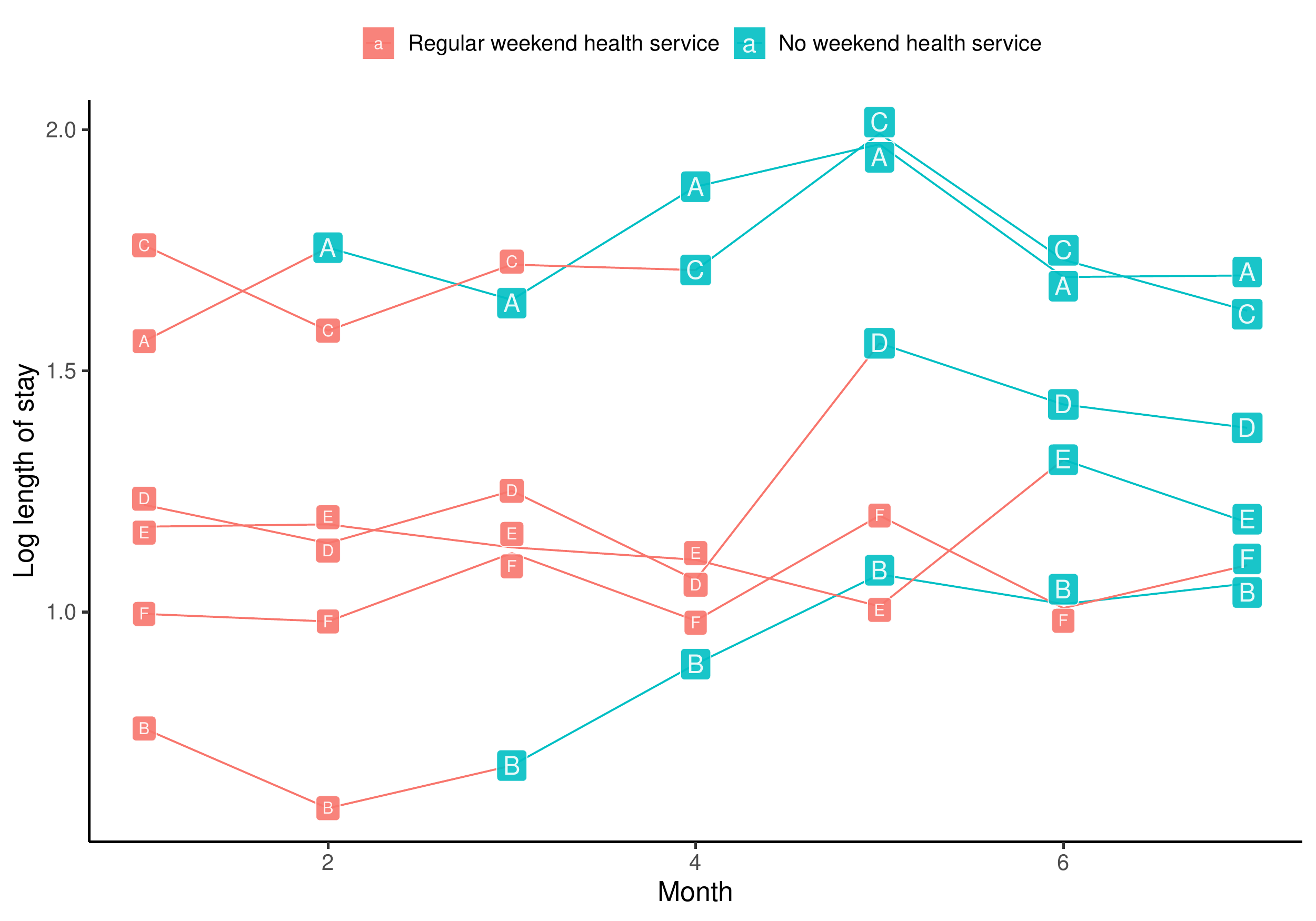}
  \caption{\footnotesize Australia weekend health services disinvestment trial:
    a 7-month stepped-wedge design with monthly ward-level mean outcomes. \qzz{}{The six hospital wards are indexed by A,B,C,D,E and F.}}
  \label{fig:ward-mean}
\end{figure}

A somewhat unusual feature of the design of this trial is that the
hospital wards received treatment (no weekend health services) in a
staggered fashion. This is often referred to as the ``stepped-wedge''
design. In the first month of the trial period, all 6
wards received regular weekend health service. In each of the
following 6 months, one ward crossed over to the treatment, and the
order was randomized at the beginning of the trial. The dataset
contains patient-level information including when and where they
were hospitalized, their length of stay, and other demographic and
medical information. More details about the data can be found in the
trial report
\citep{haines17_impac_disin_from_weeken_allied}. \Cref{fig:ward-mean}
illustrates the stepped-wedge design and shows the mean outcome of
each ward in each calendar month;
the average (log-transformed) length of stay tends to be higher after the
treatment, but more careful analysis is required to decide if such a
pattern is statistically meaningful in some way.

\subsection{\qzz{}{Trial analysis via (quasi-)randomization tests}}

We say a patient is \emph{exposed} to the treatment if there is no
weekend health services when the patient was admitted to a hospital
ward. The exposure status is jointly determined by the actual
treatment (crossover order of the wards) and when and where the
patient was admitted. This motivates seven
permutation tests for the sharp null hypothesis that removing the
weekend allied health services has no effect on the length of
stay. These tests differ in which variable(s) they
permuted. In particular, we considered permuting \qzz{ the crossover order
of the hospital wards, calendar months during which the patients
visited the hospital, and/or hospital wards visited by the
patient.}{
\begin{enumerate}
        \item Crossover: the crossover order of the hospital wards to be exposed to the treatment;
        \item Time: calendar months during which the patients visited the hospital;
        \item Ward: hospital wards visited by the patient.
\end{enumerate}
} As the trial only randomizes the crossover order, only \qzz{one
permutation test}{the test permuting crossover} qualifies as a randomization test according to our
definition in the Introduction. All six other tests that
involve permuting other variables are instances of quasi-randomization
tests in our terminology \qzz{}{because their permuted variables are
  not randomized in the trial. The quasi-randomization tests may have
  smaller p-values than the randomization test, but their validity
  requires the exchangeability assumption which may not hold. For
  instance, it is questionable to permute the
  admission times if the patients have seasonal diseases or permute
  the hospital wards if the wards have different specialties (which is
  indeed the case in this trial).}

We considered three test statistics for the permutation tests. The first
statistic $T_1$ is simply the exposed-minus-control difference in the mean
outcome; equivalently, this can be obtained by the least-squares
estimator for a simple linear regression of log length of stay on
exposure status (with intercept). The second statistic $T_2$ is
the estimated coefficient of the exposure status in the linear regression that
adjusts for the hospital wards, and the third statistic $T_3$ further
adjusts for the time of hospitalization (in calendar month).

\begin{table}[t]
\renewcommand{\arraystretch}{0.6}
\begin{center}
  \caption{\footnotesize         Results of (quasi-)randomization
  tests applied to the Australia weekend health services
  disinvestment trial.  $T_1$ is the exposed-minus-control difference of log length of stay.
  $T_2$ is the coefficient of exposure
 status in the linear regression of log length of stay on treatment status
 and hospital ward.  $T_3$ is the coefficient of exposure
 status in the linear regression of log length of stay on exposure status,
 hospital ward, and time of hospitalization. Results of the permutation tests
 are compared with the output of the corresponding linear models
 assuming normal homoskedastic noise. (CI: 90\% Confidence Interval.)}
  \label{tab:australia}
\resizebox{15cm}{!}{
\begin{tabular}{lcccccc}
  \toprule
  & \multicolumn{2}{c}{$T_1$ (adjust for nothing)} &
                                                     \multicolumn{2}{c}{$T_2$
                                                     (adjust for
                                                     ward)} &
                                                              \multicolumn{2}{c}{$T_3$
                                                              (adjust
                                                              for ward
                                                              \& time)} \\
    & p-value & CI & p-value & CI & p-value & CI \\
  \midrule
\emph{Randomization test}  & $0.0833$ & [-0.09, 0.72] & $0.0042$ & [0.06, 0.3] & $0.0069$ & [0.06, 0.31] \\
  \midrule
  \emph{Quasi-Randomization tests} & & & & \\
Time  & $0.0000$ & [0.13, 0.24] & $0.0000$ & [0.11, 0.21] & $0.0000$ & [0.09, 0.23] \\
Ward  & $0.0000$ & [0.30, 0.42] & $0.0049$ & [0.04, 0.19] & $0.0000$ & [0.10, 0.25] \\
Time \&
ward  & $0.0000$ & [0.25, 0.33] & $0.0000$ & [0.11, 0.22] & $0.0000$ & [0.09, 0.23] \\
Crossover
\& time  & $0.0029$ & [0.08, 0.47] & $0.0000$ & [0.12, 0.21] & $0.0000$ & [0.09, 0.23] \\
Crossover
\& ward  & $0.0000$ & [0.29, 0.41] & $0.0093$ & [0.02, 0.18] & $0.0001$ & [0.08, 0.24] \\
Crossover,
time \&
ward  & $0.0000$ & [0.24, 0.33] & $0.0000$ & [0.11, 0.21] & $0.0001$ &
                                                                       [0.09, 0.23] \\
  \midrule
  \emph{Linear model} & $0.0000$ & [0.24, 0.34] & $0.0000$ & [0.11, 0.22] & $0.0001$ & [0.08, 0.24] \\
  \bottomrule
\end{tabular}}
\end{center}
\end{table}

\subsection{\qzz{}{Results: Randomization test v.s. Quasi-randomization tests}}

\Cref{tab:australia} shows the one-sided p-values (alternative
hypothesis is positive treatment effect) of the seven permutation
tests with these three statistics. Confidence intervals were obtained
by inverting the two one-sided permutation tests of null hypotheses
with varying constant treatment effects (see
\Cref{subsect:partial_sharp_null} for more details).
Results are further compared with the corresponding output of the two
linear models assuming normal homoskedastic noise.
It may be useful to know that the adjusted $R^2$ of the first linear model is only 1.8\%,
while the adjusted $R^2$ of the second and third models are both about 8.4\%.

The randomization test that only permutes the crossover order almost
always gave the largest $p$-value and widest confidence
intervals. This is perhaps not too surprising given that there are
only $6!=720$ permutations in total and thus the smallest possible
permutation p-value is $1/720 \approx 0.0014$. Using better statistics
($T_2$ or $T_3$) substantially reduces the length of the confidence intervals.
On the other hand, the quasi-randomization tests, \qzz{}{while generally
  giving smaller $p$-values}, were quite sensitive to the choice of the
test statistic. In particular, the four
quasi-randomization tests that permute ward (and other
variables) produced non-overlapping confidence intervals with $T_1$
and $T_2$. The same phenomenon also occurs with the normal linear
model (last row of the table).

\begin{figure}[h]
  \centering
  \includegraphics[width = 0.75\textwidth]{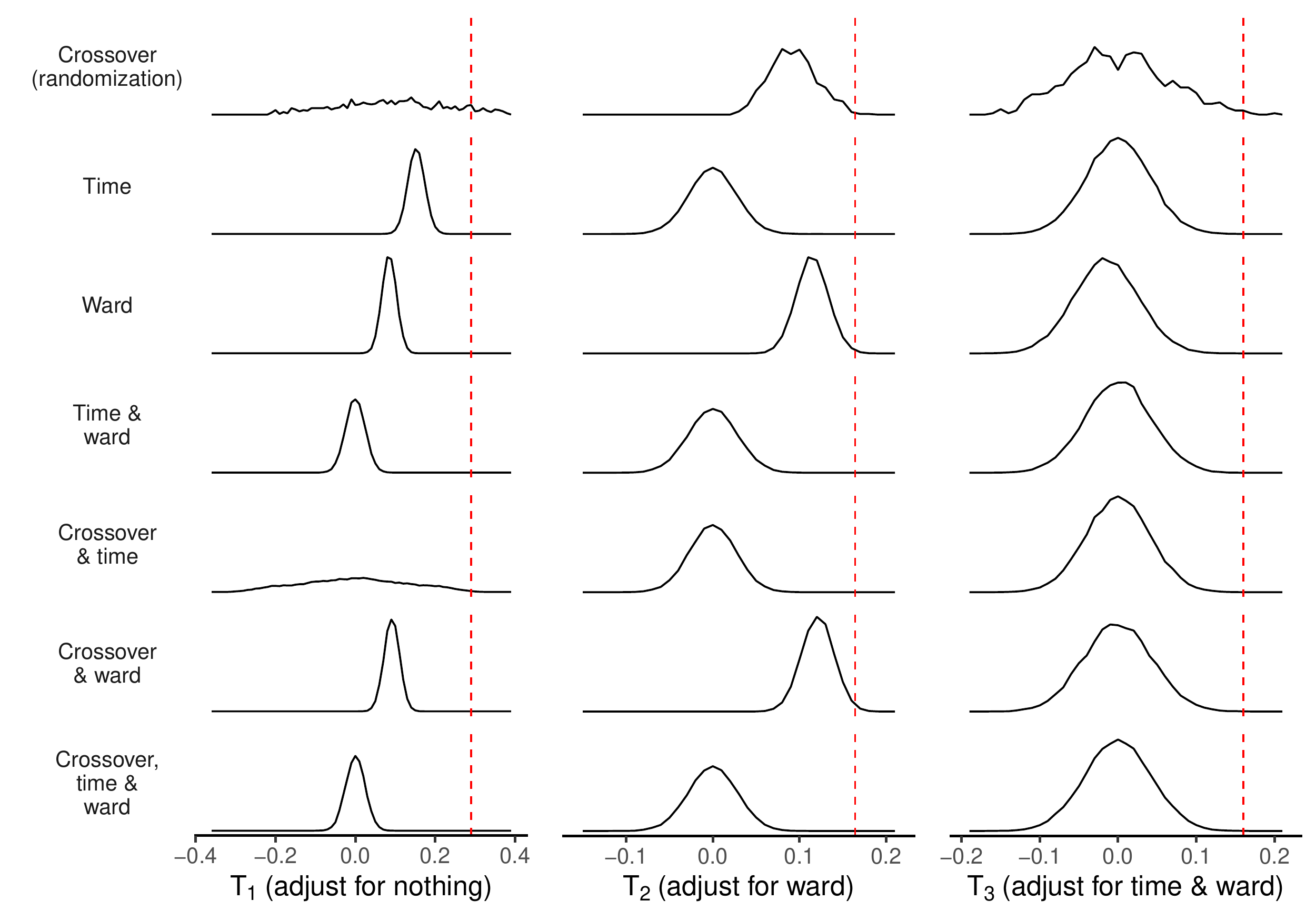}
  \caption{\footnotesize Australia weekend health services disinvestment trial:
    (quasi-)randomization distributions of three different test statistics.}
  \label{fig:random-dist}
\end{figure}

To better understand the difference between randomization and
quasi-randomization tests, \Cref{fig:random-dist} shows the
randomization distributions of $T_1$, $T_2$, and $T_3$ when different
variables are permuted.  The quasi-randomization distributions are
approximated by Monte Carlo simulation and the kernel density
estimator. \qzz{}{Each red dash line in the figure indicates the test
  statistics on the observed data without any permutation. Each
  p-value in \Cref{tab:australia} is calculated by computing the
  corresponding area of the (quasi)-randomization distribution---the
  black curve---above the observed statistic---the dashed red line. A
  small p-value means that the observed test statistics is more extreme than the
  test statistics on the permuted
  data for most of the permutations considered, which is a strong
  evidence of a positive treatment effect on the log length of stay
  conjectured in \Cref{fig:ward-mean}.
}

Notably, in the top row \qzz{}{of \Cref{fig:random-dist}},
the randomization distribution of $T_1$ is quite flat, indicating low
power. In contrast, the distributions of $T_2$ and $T_3$ have sharper
peaks. Interestingly, the randomization distribution of $T_2$
\qzz{}{(adjusted for ward)} is clearly not centred at $0$.
This is due to a general upward trend in the length of stay
\qzz{}{over the course of this trial} as shown in \Cref{fig:ward-mean}.
\qzz{}{Since the wards gradually crossed over to the treatment group}, this
\qzz{result is expected if the treatment indeed has a
positive causal effect}{trend confounds the causal effect under
investigation}.  Thus, when the crossover order and/or hospital ward
are permuted, the exposure status would have a positive coefficient in
the linear model that does not adjust for time.

\qzz{}{In the other rows of \Cref{fig:random-dist}, the distributions
  of $T_1$ and $T_2$ are centred at different places depending on
  which variables are permuted.  This explains why inverting the
  quasi-randomization tests based on $T_1$ and $T_2$ gives
  non-overlapping confidence intervals in
  \Cref{tab:australia}. Because $T_3$ adjusts both time and ward, its
  permutation distributions are much less affected by permuting time
  and ward, so the result of the quasi-randomization tests is very
  close to the randomization test (top row in that column) which only
  permutes the crossover order.
}

\subsection{\qzz{}{Recap}}

\qzz{In conclusion, the nature and validity of a permutation test clearly
depend}{The real data example gives a clear demonstration of how a
permutation test} on the randomness it tries to exploit. The
distinction between the randomization and (quasi-)randomization tests
\qzz{in the above example is practically relevant}{lead to practically
different conclusions}. \qzz{In this example, only
the test that solely permutes the crossover order is a randomization
test.}{} Using a randomization test protects against model misspecification and allows us
to take advantage of a better model \qzz{in a safe manner}{safely in the test statistics} \qzz{}{without sacrificing validity or marking additional assumption.} In contrast,
quasi-randomization tests and tests based on the normal linear
model are sensitive to model specification and \qzz{, with a reasonably good model,}{} tend to overstate statistical significance. \qzz{To interpret those
quasi-randomized p-values, we need to justify the exchangeability of
the permuted variables. In this example, exchangeability of admission
time and ward is difficult to defend. For example, the 6 hospital
wards in this trial have different specialities, so a patient admitted
for orthopaedic surgery in ward A may be quite different from a
patient admitted for stroke in ward B.}


%% file: 3_Single_CRT_revised.tex
Next, we provide a \qzz{review}{general framework} of (conditional) randomization
tests \qzz{from a causal inference perspective}{by formalizing and
generalizing what has become a ``folklore'' in causal
inference after the strong advocation by \citet{rubin80comment} and
\citet{rosenbaum02_covar_adjus_random_exper_obser_studies}}. Many
quasi-randomization tests in
the literature (such as independence testing and conformal prediction)
do not test a causal hypothesis but still fall within our
framework \qzz{}{by conceiving imaginary randomization (e.g.\ through
an i.i.d.\ or exchangeability
assumption)}. \qzz{Nevertheless,}{Importantly,} it is often
helpful to construct artificial ``potential outcomes'' in those problems to
fully understand the underlying assumption; see \Cref{sec:examples}
for some examples.


\subsection{Potential outcomes and randomization}
\label{sec:potent-outc-fram}

Consider an experiment on $N$ units in which a treatment variable $\bm
Z \in \mathcal{Z}$ is randomized. We use boldface $\bm Z$ to emphasize that the
treatment $\bm Z$ is usually multivariate. Most experiments assume
that $\bm Z = (Z_1,\dotsc,Z_N)$ collects a common attribute of the
experimental units (e.g., whether a drug is administered). However,
this is not always the case and the dimension of the treatment
variable $\bm Z$ is not important in the general theory. For example, in
the Australia weekend health services disinvestment trial in
\Cref{sec:example}, the treatment $\bm Z$ (crossover order) is
randomized at the ward level, while the patient level outcomes (length of stay). So $\mathcal{Z}$ has $6!=720$
permutations of the wards.
What the theory  below requires is that
(i) $\bm Z$ is randomized in an exogenous way by the experimenter; 
  random number generator)
(ii) the distribution of $\bm Z$ is known \qzz{( often}{(often} called the
  \emph{treatment assignment mechanism});
and (iii) one can reasonably define or conceptualize the potential
  outcomes of the experimental units under different treatment
  assignments.

To formalize these requirements, we adopt the potential
outcome (also called the Neyman-Rubin or counterfactual) framework
for causal inference
\citep{holland1986statistics,neyman1923application,rubin1974estimating}. In
this framework, unit
$i$ has a vector of real-valued \emph{potential outcomes} (or
\emph{counterfactual outcomes}) $(Y_i(\bm z)\mid
\bm z\in \mathcal{Z})$. We assume the \emph{observed outcome} (or
\emph{factual outcome}) for unit $i$ is given by $Y_i = Y_i(\bm Z)$,
where $\bm Z$ is
the realized treatment assignment. This is often referred to as the
\emph{consistency assumption} in the causal inference literature. In
our running example, $Y_i(\bm z)$ is the (potential) length of
stay of patient $i$ had the crossover order of the wards been $\bm z$.
When the treatment $\bm Z$ is an $N$-vector, the
\emph{no interference} assumption is often invoked to reduce the
number of potential outcomes; this essentially says that $Y_i(\bm z)$
only depends on $\bm z$ through $z_i$.\footnote{The
  well-known \emph{stable unit treatment value assumption} (SUTVA)
  assumes both no interference and consistency
  \citep{rubin80comment}.}
However, our theory does not rely on this assumption but treats it as part of the sharp null hypothesis
introduced below. 

It is convenient to introduce some vector notation for the potential
and realized outcomes. Let $\bm Y(\bm z) = (Y_1(\bm z), \dotsc,
Y_N(\bm z)) \in \mathcal{Y} \subseteq \mathbb{R}^N$ and $\bm Y =
(Y_1,\dotsc,Y_N) \in \mathcal{Y}$. Furthermore, let $\bm W = (\bm Y
(\bm z): \bm z \in \mathcal{Z}) \in \mathcal{W}$ collect all the
potential outcomes (which are random variables defined on the same
probability space as $\bm Z$). We will call $\bm W$ the \emph{potential
  outcomes schedule}, following the terminology in
\citet{freedman2009statistical}.\footnote{Freedman actually
  called this \emph{response schedule}.} This is also known as the
\emph{science table} in the literature \citep{rubin2005causal}. It
may be helpful to view potential outcomes $\bm Y(\bm z)$ as a
(vector-valued) function from $\mathcal{Z}$ to $\mathcal{Y}$; in
this sense, $\mathcal{W}$ consists of all functions from
$\mathcal{Z}$ to $\mathcal{Y}$.

Using this notation, The following assumption formally defines a
randomized experiment.
\begin{assumption}[Randomized experiment] \label{assump:randomization}
  $\bm Z \independent \bm W$ and the density function $\pi(\cdot)$ of
  $\bm Z$ (with respect to some reference measure on
  $\mathcal{Z}$) is known and positive everywhere. 
\end{assumption}

We write the conditional distribution of $\bm Z$ given $\bm W$ in
\Cref{assump:randomization} as $\bm Z \mid \bm W \sim
\pi(\cdot)$. This assumption formalizes the requirement that $\bm Z$ is
randomized in an exogenous way. Intuitively, the potential outcomes schedule
$\bm W$ is determined by the nature of experimental units. Since $\bm
Z$ is randomized by the experimenter, it is reasonable to assume that
$\bm Z \independent \bm W$.
In many experiments, the treatment is randomized according to some
other observed covariates $\bm X$ (e.g., characteristics of the units
or some observed network structure on the units). This can be dealt
with by assuming $\bm Z \independent \bm W \mid \bm X$ in
\Cref{assump:randomization} instead. Notice that in this case the
treatment assignment mechanism $\pi$ may depend on $\bm X$. But to
simplify the exposition, unless otherwise mentioned we will simply treat $\bm X$ as
fixed, so $\bm Z \independent \bm W$ is still true (in the
conditional probability space with $\bm X$ fixed at the observed value).


For the rest of this article, we will assume $\bm Z$ is discrete so
$\mathcal{Z}$ (e.g., $\mathcal{Z} = \{0,1\}^{N}$) is finite.


\subsection{Partially sharp null hypotheses}\label{subsect:partial_sharp_null}

As \citet{holland1986statistics} pointed out, the fundamental problem
in causal inference is that only one potential outcome can be observed
for each unit under the consistency assumption. To overcome this
problem, additional assumptions on $ \bm W$ beyond
randomization (\Cref{assump:randomization}) must be placed. In
randomization inference, the required additional assumptions are
(partially) sharp null hypotheses that relate different potential
outcomes.\footnote{In super-population inference, the fundamental
  problem of causal
  inference is usually dealt with by additional distributional
  assumptions, such as the potential outcomes of different units are
  independent and identically distributed.}

A typical (partially) sharp null hypothesis assumes that certain
potential outcomes are equal or related in certain ways. As a concrete
example, the no interference assumption
assumes that $Y_i(\bm z) = Y_i(\bm z^{*})$ whenever $z_i = z_i^{*}$. This
allows one to simplify the notation $Y_i(\bm z)$ as $Y_i(z_i)$. The no
treatment effect hypothesis (often referred to as Fisher's sharp or
exact null hypothesis) further assumes that $Y_i(z_i) = Y_i(z_i^{*})$
for all $i$ and $z_i,z_i^{*}$. When the treatment of each unit is
binary (i.e., $Z_i$ is either $0$ or $1$), under the no
interference assumption we
may also consider the null hypothesis that the treatment effect is
equal to a constant $\tau$, that is, $H_0: Y_i(1) - Y_i(0) = \tau$ for all
$i$. Under the consistency assumption, this allows us to impute the
potential outcomes as $Y_i(z_i) = Y_i + (z_i - Z_i) \tau$ for $z_i =
0, 1$.

More abstractly, a (partially) sharp null hypothesis $H$ defines
a number of relationships between the potential outcomes. Each
relationship allows us to impute some of the potential outcomes if
another potential outcome is observed (through consistency). We
can summarize these relationships using a set-valued mapping:

\begin{definition} \label{def:hypothesis}
  A partially sharp null hypothesis defines an \emph{imputability mapping}
$  \mathcal H:~ \mathcal{Z} \times \mathcal{Z} \to 2^{[N]}, (\bm  z, \bm z^{*}) \mapsto \mathcal{H}(\bm  z, \bm
                   z^{*}),$
  where $\mathcal{H}(\bm z, \bm z^{*})$ is the largest subset of $[N]$
  such that $\bm Y_{\mathcal{H}(\bm z, \bm z^{*})}(\bm z^{*})$ is
  imputable from $\bm Y(\bm z)$ under $H$.
\end{definition}

If we assume no interference and no treatment effect, all the potential
outcomes are observed or imputable regardless of the realized
treatment assignment $\bm z$, so
$\mathcal{H}(\bm z, \bm z^{*}) = [N]$. In this case, we call $H$ a
\emph{fully sharp} null hypothesis. In more sophisticated problems,
$\mathcal{H}(\bm z, \bm z^{*})$ may depend on $\bm z$ and $\bm z^{*}$
in a nontrivial way and we call such hypothesis \emph{partially
  sharp}. The concept of imputability has appeared before in
\citet{basse2019randomization} and \citet{puelz2019graph}, though
imputability was tied to test statistics \qzz{}{under a hypothesis
(see \Cref{def:stat-imputable} below) instead of the hypothesis
itself}.

In the Australia trial example (\Cref{sec:example}), we made an
implicit ``no interference'' type assumption when obtaining the
confidence intervals---we assumed that $Y_i(\bm z)$ only
depends on the implied binary exposure status $D_i(\bm z)$ of the $i$th
patient by the crossover order $\bm z$. That is, given when and where
a patient is admitted, the patient's potential outcome only depends on
whether that ward has already crossed over to the treatment group
according to $\bm z$. This would be
violated when there is interference between the wards or the effect of
ending the weekend health services is time-varying.
This assumption allows us to abbreviate
$Y_i(\bm z)$ as $Y_i(D_i(\bm z))$ and impute the potential outcome
$Y_i(0)$  under $H_0: Y_i(1) - Y_i(0) =
\tau$. In the permutation tests, test statistics were
computed using the three linear models in
\Cref{sec:example} with the shifted outcomes $Y_i-D_i(\bm Z)\tau,~
i=1,\dotsc,N$. The permutation tests were subsequently inverted to
obtain confidence intervals of $\tau$.

\subsection{Conditional randomization tests (CRTs)}

\label{sec:crt-discrete}

To test a (partially) sharp null hypothesis, a randomization test compares an
observed test statistic with its \emph{randomization distribution},
which is given by the value of the statistic under a random treatment
assignment. However, it may be impossible to compute the entire
randomization distribution when some potential outcomes are not
imputable (i.e.\ when $\mathcal{H}(\bm z, \bm z^{*})$ is smaller
than $[N]$). To tackle this issue, we confine ourselves to a
smaller set of treatment assignments. This is formalized in the next
definition.

\begin{definition}\label{def:CRT}
  A \emph{conditional randomization test} (CRT) for a treatment
  $\bm Z$ is defined by
(i) \label{item:crt_1} A \emph{partition} $\mathcal{R} = \{\mathcal{S}_m\}_{m=1}^M$ of $\mathcal{Z}$ such that
    $\mathcal{S}_1,\dotsc,\mathcal{S}_M$ are disjoint subsets of
    $\mathcal{Z}$ satisfying $\mathcal{Z} = \bigcup_{m=1}^M
    \mathcal{S}_{m}$; and
(ii) A collection of \emph{test statistics}
    $(T_m(\cdot,\cdot))_{m=1}^M$, where
    $T_m:\mathcal{Z} \times \mathcal{W} \to \mathbb{R}$ is a
    real-valued function that computes a test statistic for each
    realization of the treatment assignment $\bm Z$ given the potential
    outcomes schedule $\bm W$.
\end{definition}

\qzz{}{Methods to construct $\mathcal{R}$ and examples will be
discussed in the following sections.}

Any partition $\mathcal{R}$ defines an equivalent relation
$\equiv_{\mathcal{R}}$ and vice versa, so
$\mathcal{S}_1,\dotsc,\mathcal{S}_M$ are simply the equivalence
classes generated by $\equiv_{\mathcal{R}}$.
With an abuse of notation, we let $\mathcal{S}_{\bm z} \in
\mathcal{R}$ denote the equivalence class containing $\bm z$. For any
$\bm z \in \mathcal{S}_m$, we thus have $\mathcal{S}_{\bm z} = \mathcal{S}_m$
and $T_{\bm z}(\cdot,\cdot) = T_m(\cdot,\cdot)$. This notation is
convenient because the p-value of the CRT defined below conditions
on $\bm Z^{*} \in \mathcal{S}_{\bm z}$ when we observe $\bm Z = \bm
z$. The following property follows immediately from the fact that
$\equiv_{\mathcal{R}}$ is an equivalence relation:

\begin{lemma}[Invariance of conditioning sets and test
  statistics] \label{lem:same-st}
  For any $\bm z \in \mathcal{Z}$ and $\bm z^{*} \in \mathcal{S}_{\bm
    z}$, we have $
  \bm z \in \mathcal{S}_{\bm z}$, $\mathcal{S}_{\bm z^{*}} =
  \mathcal{S}_{\bm z}$ and $ T_{\bm z^{*}}(\cdot, \cdot) = T_{\bm
    z}(\cdot, \cdot)$.
\end{lemma}

\begin{definition} \label{def:p-value}
  The \emph{p-value} of the CRT in \Cref{def:CRT} is given by
  \begin{equation}
    \label{eq:p-value}
    P(\bm Z, \bm W) = \mathbb{P}^{*} \{T_{\bm Z}(\bm Z^{*}, \bm W)
    \leq T_{\bm Z}(\bm Z, \bm W) \mid \bm Z^{*}\in \mathcal{S}_{\bm
      Z}, \bm Z, \bm W \},
  \end{equation}
  where $\bm Z^{*}$ is an independent copy of $\bm Z$ conditional on
  $\bm W$ and the notation $\mathbb{P}^{*}$ is used to emphasize that
  the probability is taken over $\bm Z^{*}$.
\end{definition}

Because $\bm Z \independent \bm W$ (\Cref{assump:randomization}), $\bm
Z^{*}$ is independent of $\bm Z$ and $\bm W$ and $\bm Z^{*} \sim
\pi(\cdot)$.
The invariance property in \Cref{lem:same-st} is important because it
ensures that, when computing the p-value, the same conditioning set is
used for all the assignments within it. By using the equivalence
relation $\equiv_{\mathcal{R}}$ defined by the partition
$\mathcal{R}$, we can rewrite \eqref{eq:p-value} as
\[
  P(\bm Z, \bm W) = \mathbb{P}^{*} \{T_{\bm Z}(\bm Z^{*}, \bm W)
  \leq T_{\bm Z}(\bm Z, \bm W) \mid \bm Z^{*} \equiv_{\mathcal{R}} \bm
  Z, \bm Z, \bm W \}.
\]
When $\mathcal S_{\bm z} = \mathcal{Z}$ for all $\bm z \in
\mathcal{Z}$
this reduces to an unconditional randomization test.

Notice that $T_{\bm Z}(\bm Z^{*}, \bm W)$ generally depends on some
unobserved potential outcomes in $\bm W$. Thus the p-value \eqref{eq:p-value} may
not be computable if the null hypothesis does not make enough
restrictions on how $T_{\bm Z}(\bm Z^{*}, \bm W)$ depends on $\bm W$. By using the imputability mapping
$\mathcal{H}(\bm z, \bm z^{*})$ in \Cref{def:hypothesis}, this is
formalized in the next definition.

\begin{definition} \label{def:stat-imputable}
  Consider a CRT defined by the partition
  $\mathcal{R} = \{\mathcal{S}_m\}_{m=1}^M$ and test statistics
  $(T_m(\cdot,\cdot))_{m=1}^M$. We say the test statistic
  $T_{\bm z}(\cdot,\cdot)$ is
  \emph{imputable} under a partially sharp null hypothesis $H$ if for
  all $\bm z^{*} \in
  \mathcal{S}_{\bm z}$, $T_{\bm z}(\bm z^{*},
  \bm W)$ only depends on
  the potential outcomes schedule $\bm W =
  (\bm Y(\bm z) : \bm z \in \mathcal{Z})$ through its imputable
  part $\bm Y_{\mathcal{H}(\bm z, \bm z^{*})}(\bm z^{*})$.
\end{definition}

\begin{lemma}\label{lemma:computable}
  Suppose \Cref{assump:randomization} is satisfied and $T_{\bm
    z}(\cdot,\cdot)$ is imputable under $H$ for all $\bm z \in
  \mathcal{Z}$. Then the p-value $P(\bm Z, \bm W)$ only depends on
  $\bm Z$ and $\bm Y$.
\end{lemma}

\begin{definition} \label{def:p-value-computable}
  Under the assumptions in \Cref{lemma:computable}, we say the p-value
  is \emph{computable} under $H$ and denote it, with an abuse of
  notation, by $P(\bm Z, \bm Y)$.
\end{definition}

Given a computable p-value, the CRT then rejects the null hypothesis
$H$ at significance level $\alpha \in [0,1]$ if $P(\bm Z, \bm Y)
\leq \alpha$. The next theorem establishes the validity of this test.

\begin{theorem}\label{thm:valid}
  Consider a CRT defined
  by the partition
  $\mathcal{R} = \{\mathcal{S}_m\}_{m=1}^M$ and test statistics
  $(T_m(\cdot,\cdot))_{m=1}^M$. Then the p-value $P(\bm Z,
  \bm W)$ \qzz{}{is valid} in the sense that it stochastically
  dominates the uniform distribution on $[0,1]$\qzz{in the following sense}{}:
  \begin{equation}
    \label{eq:stoc-domi}
    \mathbb{P}\left\{ P(\bm Z, \bm W)\leq \alpha \mid \bm Z \in
      \mathcal{S}_{\bm z}, \bm
      W\right\}\leq \alpha, \quad \forall \alpha\in[0,1], \bm z \in
    \mathcal{Z}.
  \end{equation}
  In consequence, given \Cref{assump:randomization}
  and a partially sharp null hypothesis $H$, if $P(\bm Z, \bm W)$ is
  computable, then
  \begin{equation}
    \label{eq:stoc-domi-marginal}
    \mathbb{P}\left\{ P(\bm Z, \bm Y)\leq \alpha \mid \bm W\right\}\leq \alpha,
    \quad \forall \alpha\in[0,1].
  \end{equation}
\end{theorem}

Note that by marginalizing \eqref{eq:stoc-domi-marginal} over the potential outcomes schedule $\bm
W$, we obtain
\[
  \mathbb{P}\left\{ P(\bm Z, \bm Y)\leq \alpha \right\}\leq \alpha,
  \quad \forall \alpha\in[0,1].
\]
The conditional statement \eqref{eq:stoc-domi-marginal} is stronger as
it means that the type I error is always controlled for the given
samples. \qzz{No assumptions about the sample are required.}{In addition to
\Cref{assump:randomization}, no assumptions are required about the sample.
}

\subsection{Nature of conditioning}
\label{sec:conditioning-sets}

To construct the partition $\mathcal{R}$, one common approach is to
condition on a function of, or more precisely, a random variable $G =
g(\bm Z)$ generated by $\bm Z$. This idea is formalized by the next result,
which immediately follows by
defining the equivalence relation $\bm z^{*} \equiv_{\mathcal{R}}
\bm z$ when $g(\bm z^{*}) = g(\bm z)$.

\begin{proposition} \label{prop:condition-on-function}
  Any function $g: \mathcal{Z} \to \mathbb{N}$ defines a countable
  collection of invariant conditioning sets
  $\mathcal{S}_{\bm z} = \{\bm z^{*} \in \mathcal{Z}: g(\bm z^{*}) = g(\bm
  z)\}$.
\end{proposition}

\qzz{}{Tests of this form have appeared before in
\citet{zheng08_multi_center_clinic_trial} and
\citet{hennessy2016conditional} to deal with covariate imbalance in
randomized experiments.
As an example, consider a Bernoulli trial with 10 females (unit
$i=1,\dotsc 10$) and
10 males (unit $i=11,\dots,20$). Let $g(\bm z) = \sum_{i=1}^{10} z_i$
denote the number of treated females in any assignment $\bm z$.
Suppose the realized randomization has has only
2 treated females, that is, $g(\bm Z) = 2$, due to chance. Using the
conditioning set
$\mathcal{S}_{\bm Z} =
\{\bm z^{*} \in \{0,1\}^{20}: g(\bm z^{*}) = 2\}$, the CRT only
compares $\bm Z$ with other assignments $\bm z^{*}$ also with 2
treated females, which removes any potential bias due to a gender
effect. In this example, $g$ maps $\mathcal{Z}
= \{0,1\}^{20}$ to $\{0,1,\dotsc,10\}$, and the partiion of
$\mathcal{Z}$ is given by $\mathcal{R}=\{\mathcal{S}_m\}_{m=0}^{10} $
where $\mathcal{S}_m = \{\bm z\in \mathcal{Z}, g(\bm z ) = m\}.$
}

\qzz{Alternatively,}{More generally,} one can consider a measure-theoretic formulation of
conditional randomization tests. For example, let $
\mathcal{G} = \sigma\left( \{ \bm Z \in \mathcal{S}_{m}\}_{m =
    1}^{\infty}\right).
$ be the $\sigma$-algebra generated by the conditioning events in
\eqref{eq:p-value}.
Because $\{\mathcal{S}_m\}_{m=1}^{\infty}$ is a partition,
$\mathcal{G}$ consists of all countable unions of
$\{\mathcal{S}_m\}_{m=1}^{\infty}$. This allows us to rewrite
\eqref{eq:stoc-domi} as
\begin{equation*}
  \label{eq:stoc-domi-sigma}
  \P\left(P(\bm Z, \bm W) \leq \alpha \mid \mathcal{G}, \bm W\right)
  \leq \alpha, \quad \forall \alpha \in [0,1].
\end{equation*}
\qzz{}{This is equivalent to
\[
  \P\left(P(\bm Z, \bm W) \leq \alpha \mid \bm Z\in \mathcal{S}_m, \bm W\right)
  \leq \alpha, \quad \forall \alpha \in [0,1] \text{ and } m=1,2,\dotsc.
\]
}
This measure-theoretic formulation is useful for extending the theory
above to continuous treatments and consider the
structure of conditioning events in multiple CRTs, which will be
considered in a separate article.


\subsection{Post-randomization}
\label{sec:randomized-crts}

In many problems, there are several ways to construct the conditioning
event/variable; see e.g.\ \Cref{sec:practical-methods}. In
such a situation, a natural idea is to post-randomize the test.

Consider a collection of CRTs defined by $\mathcal{R}(v) =
\{\mathcal{S}_m(v)\}_{m=1}^{M}$ and $
(T_m(\cdot,\cdot;v))_{m=1}^{M}$ that are indexed by
$v \in \mathcal{V}$ where $\mathcal{V}$ is countable. In the example
of bipartite graph representation \qzz{}{introduced below} in
\Cref{sec:bipartite-cliques}, $v$ can be a biclique decomposition of
the graph. Each $v$ defines a p-value
\begin{equation*}
  \label{eq:p-value-random}
  P(\bm Z, \bm W; v) = \mathbb{P}^{*} \{T_{\bm Z}(\bm Z^{*}, \bm W;v)
  \leq T_{\bm Z}(\bm Z, \bm W;v) \mid \bm Z^{*}\in
  \mathcal{S}_{\bm Z}(v), \bm W \},
\end{equation*}
where $\bm Z^{*}$ is an independent copy of $\bm Z$, and we may use a
random value $P(\bm Z, \bm W, V)$ where $V$ is drawn by the analyst
and thus independent of $(\bm Z, \bm W)$. It immediately follows from
\Cref{thm:valid} that this defines a valid test in the sense that
\begin{equation} \label{eq:valid}
  \mathbb{P}\left\{ P(\bm Z, \bm W; V)\leq \alpha \mid \bm Z \in
    \mathcal{S}_{\bm z}(V), \bm
    W, V\right\}\leq \alpha, \quad \forall \alpha\in[0,1], \bm z \in
  \mathcal{Z}.
\end{equation}

A more general viewpoint is that we may
condition on a random variable $G = g(\bm Z, V)$ that depends on not only
the randomness introduced by the experimenter in $\bm Z$ but also the
randomness introduced by the analyst in
$V$. \Cref{prop:condition-on-function} can then be generalized in a
straightforward way. The construction below is inspired by
\citet{bates2020causal} (who were concerned with genetic mapping)
and personal communiations with Stephen Bates.

As above, suppose $V \independent (\bm Z, \bm W)$ and $G$ has a
countable support.  Because $G$ is generated by $\bm Z$, the
conditional distribution of $G$ given $\bm Z$ is known.
Let $\pi(\cdot \mid g)$ be the density function of $\bm Z$ given
$G = g$, which can be obtained from Bayes' formula:
\[
  \pi(\bm z \mid g) = \frac{\P(G = g \mid \bm Z = \bm
    z)\pi(\bm z) }{\int \P(G = g \mid \bm Z = \bm z) \pi(\bm z)  \diff \mu(\bm z)},
\]
where $\pi$ is the density of $\bm Z$ with respect to some reference
measure $\mu$ on $\mathcal{Z}$. Let $T_g(\cdot,
\bm W)$ be the test statistic that is now indexed by
$g$ in the support of $G$. The post-randomized p-value is then
defined as
\begin{align*}
  P(\bm Z, \bm W; G) &= \mathbb{P}^{*} \left\{T_G(\bm Z^{*}, \bm W)
                       \leq T_G(\bm Z, \bm W) \mid G, \bm W, \bm Z
                       \right\},
\end{align*}
where the probability is taken over $\bm Z^{*} \mid G, \bm W
\overset{d}{=} \bm Z \mid G, \bm W$. In other words, $\bm Z^{*} \sim
\pi(\cdot \mid G)$ and the randomized p-value can be written as
\[
  P(\bm Z, \bm W; G) = \int 1_{\{T_G(\bm z^{*}, \bm W) \leq T_G(\bm Z, \bm
    W)\}} \pi(\bm z^{*} \mid G) \diff \mu(\bm z^{*}).
\]
Similar to above, we say $P(\bm Z, \bm W; g)$ is computable
if it is a function of $\bm Z$ and $\bm Y$ under the null hypothesis
and write it as $P(\bm Z, \bm Y; g)$.

\begin{theorem} \label{cor:valid-2}
  Under the setting above, the randomized CRT is valid in the
  following sense
  \begin{equation*}
    \mathbb{P}\left\{ P(\bm Z, \bm W; G)\leq \alpha \mid \bm
      W, G\right\}\leq \alpha, \quad \forall \alpha\in[0,1].
  \end{equation*}
\end{theorem}

\Cref{cor:valid-2} generalizes several results above. \Cref{thm:valid}
is essentially a special case where $G = \mathcal{S}_{\bm Z}$ is a
set. \Cref{prop:condition-on-function} is also a special case
where $G = g(\bm Z)$ is not randomized. Finally, equation \eqref{eq:valid}
amounts to conditioning on the post-randomized set $G = \mathcal{S}_{\bm
  Z}(V)$. \Cref{cor:valid-2} also generalizes a similar theorem in
\citet[Theorem 1]{basse2019randomization} by
allowing post-randomization and not
requiring imputability of the test statistic. In other words,
imputability only affects whether the p-value can be computed using
the observed data and is not necessary for the validity of the
p-value.

\qzz{}{An alternative to a single post-randomized test is to average
the p-values from different realizations of $G$. It may be shown that
the average of those p-value is valid up to a factor 2, in the sense
that the type I error is upper bounded by $2 \alpha$ if the null is
rejected when the average p-value is less than $\alpha$
\citep{ruschendorf1982random,vovk2020combining}. This strategy may be
useful when post-randomization gives rise to a large variance.}



%% file: 4_Practical_Methods.tex
This section summarizes some practical methods to construct computable
and powerful tests from the causal interference literature
\citep{aronow2017estimating,athey2018exact,basse2019randomization,bowers2013reasoning,hudgens2008towards,li2019randomization,puelz2019graph}. \qzz{}{The
following example provides some context.}

\qzz{}{
Consider an experiment that displays an advertisement (or nothing) to
the users of a social network, and we would like to test if displaying
the advertisement to a user has a spillover effect on their
friends. Each user thus has one of three exposures: directly see the
advertisement (``treated''), has a friend who sees the advertisement
(``spillover''), or has no direct or indirect exposure to the
advertisement (``control''). As the null hypothesis of no
spillover effect only relates the potential outcomes under spillover
and control, the outcomes of the treated users (denote the collection
of them by
$\mathcal{I}_{\bm Z}\subset [N]$) do not provide any information about
the hypothesis; in other words, $\mathcal{I}_{\bm Z}\cap
\mathcal{H}(\bm Z, \bm z^{*}) = \emptyset$ for all $\bm z^{*}$. This
means that a test statistic $T(\bm Z, \bm W)$ is imputable only if it
does not depend on the potential outcomes of the users in the set
$\mathcal{I}_{\bm Z}$ changing with $\bm Z$. This makes it difficult
to construct an imputable test statistic. 
}

\subsection{\qzz{}{Intersection method}}

Often, the test statistic of a CRT only depends on the
potential outcomes corresponding to the (counterfactual) treatment
and takes the form $T_{\bm z}(\bm
z^{*}, \bm W) = T_{\bm z}(\bm z^{*}, \bm Y(\bm z^{*}))$. Then the
fundamental challenge is that only a sub-vector $\bm
Y_{\mathcal{H}(\bm z, \bm z^{*})}(\bm z^{*})$ of $\bm Y(\bm z^{*})$ is
imputable under $H$. A natural idea is to only
use the imputable potential outcomes.

\begin{proposition} \label{prop:choose-h}
  Given any partition $\mathcal{R} = \{\mathcal{S}_m\}_{m=1}^{M}$
  of $\mathcal{Z}$, let $\mathcal{H}_m = \bigcap_{\bm z, \bm z^{*} \in \mathcal{S}^m}
  \mathcal{H}(\bm z, \bm z^{*})$. Then, under
  \Cref{assump:randomization}, the partition $\mathcal{R}$ and
  test statistics $(T_m(\bm z,\bm Y_{\mathcal{H}_m}(\bm
  z)))_{m=1}^{M}$ define a computable p-value.
\end{proposition}

However, the CRT in \Cref{prop:choose-h} would be powerless if
$\mathcal{H}_m$ is an empty set. More generally, the power of the CRT
depends on the size of $\mathcal{S}_m$ and $\mathcal{H}_m$, and there
is an important trade-off: with a coarser $\mathcal{R}$, the CRT is
able to utilize a larger subset
$\mathcal{S}_m$ of treatment assignments but a smaller subset
$\mathcal{H}_m$ of experimental units.
In many problems, choosing a good partition $\mathcal{R}$ is
nontrivial. In such cases, it may be helpful to impose some structure
on the imputability mapping $\mathcal{H}(\bm z, \bm z^{*})$.

\begin{definition} \label{def:level-set}
A partially sharp null hypothesis $H$ is said to have a
\emph{level-set structure} \qzz{if there exist}{with respect to a
  collection of} \emph{exposure functions}
$D_i: \mathcal{Z} \to \mathcal{D}, i=1,\dotsc,N$, if
$\mathcal{D}$ is countable and
\begin{equation}
  \label{eq:h-level-set}
\mathcal{H}(\bm z, \bm z^{*}) = \{i \in [N]: D_i(\bm z) = D_i(\bm z^{*}) \}.
\end{equation}
\end{definition}

The imputability mapping is then defined by the level sets
of the exposure functions. This would occur if, for example, the null
hypothesis only specifies the treatment effect between two exposure
levels\qzz{}{, as in our social network advertisement example}. 
\Cref{def:level-set} is inspired by \citet[Definition
3]{athey2018exact}, but the concept of
exposure mapping can be traced back to
\citet{aronow2017estimating,manski2013identification,ugander2013graph}.

An immediate consequence of the level-set structure is
that $\mathcal{H}(\bm z, \bm z^{*})$ is symmetric. Moreover, by using
the level-set structure, we can write $\mathcal{H}_m$ in
\Cref{prop:choose-h} as
\begin{equation}
  \label{eq:h-m}
  \mathcal{H}_m = \bigcap_{\bm z, \bm z^{*} \in \mathcal{S}_m}
  \mathcal{H}(\bm z, \bm z^{*}) = \left\{i \in [N]: D_i(\bm z) \text{
      is a constant
  over } \bm z \in \mathcal{S}_m\right\}.
\end{equation}
This provides a way to choose the test statistic
once the partition $\mathcal{R} = \{\mathcal S_m\}_{m=1}^{M}$ is given.

\subsection{\qzz{}{Focal units}}
\label{sec:focal-units}

We may also proceed in the other direction and choose the experimental
units first. \citet{aronow2012general} and \citet{athey2018exact}
proposed to choose a partition $\mathcal{R} =
\{\mathcal{S}_m\}_{m=1}^{M}$ such that $\mathcal{H}_m$ is equal to a
fixed subset of ``focal units'', $\mathcal{I} \subseteq [N]$, for all
$m$. Given any $\mathcal{I} \subseteq [N]$, the conditioning set is
given by all the treatment assignments such that all the units in
$\mathcal{I}$ receive the same exposure. That is,
\begin{equation}
  \label{eq:level-set-s}
  \mathcal{S}_{\bm z} = \{\bm z^{*} \in \mathcal{Z}: \mathcal{I}
  \subseteq \mathcal{H}(\bm z, \bm z^{*})\} = \{ \bm z^{*} \in
  \mathcal{Z}: \bm D_{\mathcal{I}}(\bm z^{*}) = \bm D_{\mathcal{I}}(\bm z)  \},
\end{equation}
where $\bm D_\mathcal{I}(\cdot) =
(D_i(\cdot))_{i \in \mathcal{I}}$. From the right-hand side of
\eqref{eq:level-set-s}, it is easy to see that $\{\mathcal{S}_{\bm z}:
\bm z \in
\mathcal{Z}\}$ satisfies \Cref{lem:same-st} and thus forms a partition
of $\mathcal{Z}$. Furthermore, $\{\mathcal{S}_{\bm z}: \bm z \in
\mathcal{Z}\}$ is countable because
$\mathcal{S}_{\bm z}$ is determined by $\bm D_\mathcal{I}(\bm z)$, a subset of the
countable set $\mathcal{D}^{\mathcal{I}}$. \qzz{}{In our social
  network advertisement example, the focal units can be a randomly
  chosen subset of users; see \citet{aronow2012general} and
  \citet{athey2018exact} for more discussion.}

The next proposition summarizes the method proposed by
\citet{athey2018exact} and immediately follows from our discussion
above.\footnote{\citet{athey2018exact} used the same
  test statistic in all
conditioning events, which is reflected in \Cref{prop:athey}. Our
construction further allows the test statistic
  $T_{\bm Z}(\bm z,\bm Y_{\mathcal{I}}(\bm z))$ to depend
  on $\bm Z$ through $\bm D_{\mathcal{I}}(\bm Z)$.}
\begin{proposition} \label{prop:athey}
Given a null
  hypothesis $H$ with a level-set structure in \cref{def:level-set}, and a set of focal units $\mathcal{I}\subseteq [N]$. Under
  \Cref{assump:randomization}, the partition $\mathcal{R}
  = \{\mathcal{S}_{\bm z}: \bm z \in \mathcal{Z}\}$ as defined in
  \eqref{eq:level-set-s} and any test statistic $T(\bm z,\bm
  Y_{\mathcal{I}}(\bm z))$ induce a computable p-value.
\end{proposition}


\subsection{Bipartite graph representation}
\label{sec:bipartite-cliques}

\citet{puelz2019graph} provided an alternative way to use the
level-set structure. They consider imputability mapping of the form
(suppose $0 \in \mathcal{D}$)\footnote{The ``null exposure graph'' in
  \citet{puelz2019graph} actually allows $D_i(\bm z)$ and $D_i(\bm
  z^{*})$ to belong to a prespecified
  subset of $\mathcal{D}$. This can be incorporated in our setup by
  redefining the exposure functions.}
\begin{equation}\label{eq:h-level-set-new}
\mathcal{H}(\bm z, \bm z^{*}) = \{i \in [N]: D_i(\bm z) = D_i(\bm
z^{*}) = 0 \},
\end{equation}
which is slightly more restrictive than
\eqref{eq:h-level-set}. \qzz{}{In the social network example,
  $\mathcal{H}(\bm z, \bm z^{*})$ is the subset of users who do not
  receive the advertisement directly in both $\bm z$ and $\bm z^*$.
}
The ``conditional focal units''
$\mathcal{H}_m$ in \eqref{eq:h-m} can then be written as
\begin{equation}
  \label{eq:h-m-0}
    \mathcal{H}_m = \{i \in [N]: D_i(\bm z) = 0, \forall \bm z \in
  \mathcal{S}_m\}.
\end{equation}
Their key insight is that the condition in \eqref{eq:h-m-0} can be
visualized using a bipartite graph with vertex
set $\mathcal{V} = [N] \cup \mathcal{Z}$ and edge set
 $ \mathcal{E}= \{(i, \bm z) \in [N] \times \mathcal{Z}:
  D_i(\bm z) = 0\}$ \qzz{}{connecting every unit $i$ with every assignment $\bm z$ satisfying that $D_i(\bm z) = 0$. }
\citet{puelz2019graph} referred to this as the \emph{null exposure
  graph} \qzz{}{$\mathcal{G}=(\mathcal{V},\mathcal{E}).$} Then by using \eqref{eq:h-m-0}, we have
\begin{proposition}
\qzz{}{The vertex subset}
 $\mathcal{V}_m = \mathcal{H}_m \cup  \mathcal{S}_m$ and \qzz{}{the edge subset} $
 \mathcal{E}_m= \{(i, \bm z) \in \mathcal{H}_m \times \mathcal{S}_m
 \}$ form a biclique (i.e., a
  complete bipartite subgraph) in $\mathcal{G}$.
\end{proposition}

\qzz{}{By definition, both $\mathcal{V}_m$ and $\mathcal{E}_m$ depend
  on $S_m$.} The challenging problem of finding a good partition of
$\mathcal{Z}$ is reduced to
finding a collection of large bicliques $\{ (\mathcal{V}_m,
\mathcal{E}_m)\}_{m=1}^M$ in the graph such that
$\{\mathcal{S}_m\}_{m=1}^M$ partitions $\mathcal{Z}$. This was called a
\emph{biclique decomposition} in \citet{puelz2019graph}. They
further described an approximate algorithm to find a biclique decomposition by
greedily removing treatment assignments in the largest biclique.




%% file: 5_Examples_revised.tex
Next, we examine some randomization and quasi-randomization tests
proposed in the literature. These examples not only demonstrate the
generality and usefulness of the theory above but also help to
clarify concepts and terminologies related to randomization tests.

\subsection{Fisher's exact test}
\label{sec:fishers-exact-test}

Fisher's exact test is perhaps the simplest (quasi-)randomization
test. In our notation, let $\bm Z \in \{0,1\}^N$ be the
treatment assignment for $N$ units and $Y_i(0) \in \{0,1\},
Y_i(1) \in \{0,1\}$ be the potential outcomes of each unit $i$
(so the no interferencec assumption is made). We are interested in
testing the hypothesis that the treatment has no effect
whatsoever. Because both the
treatment $Z_i$ and outcome $Y_i = Z_i Y_i(1) + (1 - Z_i) Y_i(0)$ are
binary, data can be summarized by a $2 \times 2$ table where $N_{zy}$
denotes the number of units with treatment $z$ and outcome $y$
for $z,y \in \{0,1\}$.
Let $N_{z\cdot} =
N_{z0} + N_{z1}$ and $N_{\cdot y} = N_{0y} + N_{1y}$ be the row and
column marginal totals, respectively.
Fisher observed that the probability of observing
$(N_{00},N_{01},N_{10},N_{11})$ given the marginal totals is given by
\[
 \frac{{N_{0\cdot} \choose N_{00}}{N_{1\cdot} \choose N_{10}}}{{N \choose N_{\cdot 0}}}
=  \frac{{N_{0\cdot} \choose N_{01}}{N_{1\cdot} \choose N_{11}}}{{N \choose N_{\cdot 1}}}
 = \frac{N_{0\cdot}!N_{1\cdot}! N_{\cdot 0}! N_{\cdot 1}!}{N_{00}!
    N_{01}! N_{10}! N_{11}! N!},
\]
which can then be used to compute an exact p-value by summing up the
probabilities of equally or more extreme tables. Notice that the
column marginal totals $N_{\cdot 1} = \sum_{i=1}^N Y_i$ and $N_{\cdot 0} =
\sum_{i=1}^N (1 - Y_i)$ are fixed under the sharp null hypothesis
$H_0: Y_i(0) = Y_i(1)$. Thus, a significance test that does not fix the
column margins cannot be a test of the sharp null. An example of this
is Barnard's test for the ``two binomial'' problem
\citep{barnard47_signif_tests_for_tables}, which was abandoned by
Barnard himself (see \cite{yates1984tests} and references therein) but
is still being applied in practice due to the impression that it is
more powerful than Fisher's.


The nature of Fisher's exact test depends on how the data is
generated. When the treatment $\bm Z$ is randomized, Fisher's exact
test is a randomization test. Whether it is a conditional or
unconditional test further depends on the treatment assignment
mechanism. If $\bm Z$ is completely randomized but the total number of
treated units $\sum_{i=1}^n Z_i = N_{1 \cdot}$ is fixed as in
the famous tea-tasting example
\citep{fisher1935design}\footnote{\citet{yates1984tests} called this
  the ``comparative trial''.}, Fisher's
exact test is an unconditional randomization
test. If
$\bm Z$ is generated from an independent Bernoulli trial, $N_{1 \cdot}$
is random and Fisher's exact test is a conditional randomization test
that conditions on $N_{1 \cdot}$. When $\bm
Z$ is not randomized, Fisher's exact test is a quasi-randomization
test. An example of this is the so-called ``two binomials'' problem
\citep{barnard47_signif_tests_for_tables}. There, conditioning on the
row marginal totals is often presented as a way to eliminate nuisance
parameters \citep[sec.\ 4.4-4.5]{lehmann2006testing}.

Significance testing for $2 \times 2$ tables is indeed a topic of
recurring discussion;
see
\citet{yates1984tests,little89_testin_equal_two_indep_binom_propor,ding16_poten_tale_two_by_two}
and referenes therein. Much of the conceptual confusion and controversy
can be avoided by distinguishing randomization tests from
quasi-randomization tests. 

\subsection{Permutation tests for treatment effect}
\label{sec:permutation-tests}

In a permutation test, the p-value is obtained by calculating all
possible values of the test statistics under all allowed permutations
of the observed data points. As argued in the Introduction, the name
``permutation test'' emphasizes the algorithmic perspective of the
statistical test and thus is not synonymous with ``randomization
test''.

In the context of testing treatment effect, a permutation test is
essentially a CRT that uses the following
conditioning sets (suppose $\bm Z$ is a vector
of length $N$)
\begin{equation}
  \label{eq:sz-permutation}
  \mathcal{S}_{\bm z} = \{(z_{g(1)},\dotsc,z_{g(N)}):\text{$g$
    is a permutation of $[N]$}\}.
\end{equation}
In view of \Cref{prop:condition-on-function}, a permutation test is a
CRT that conditions on the order statistics of $\bm Z$.
In permutation tests, the treatment assignments are typically assumed
to be exchangeable,
  $(Z_1,\dotsc,Z_N) \overset{d}{=} (Z_{g(1)}, \dotsc,
  Z_{g(N)})~\text{for all permutations $g$ of $[N]$},$
so each permutation of $\bm Z$ has the same probability of being
realized under the treatment assignment mechanism
$\pi(\cdot)$. See
\cite{kalbfleisch78_likel_method_nonpar_tests} for an alternative
formulation of rank-based tests based on marginal and conditional
likelihoods. 
Exchangeability makes it straightforward to compute the
p-value \eqref{eq:p-value}, as $\bm Z^{*}$ is uniformly distributed
over $\mathcal{S}_{\bm Z}$ if $\bm Z$ has
distinct elements. In this sense, our assumption that the assignment
distribution of $\bm Z$ is known (\Cref{assump:randomization}) is more
general than exchangeability. See \cite{roach2018permutation} for
some recent development on generalized permutation tests in
non-exchangeable models.

Notice that in permutation tests, the invariance of $\mathcal{S}_{\bm
  z}$ in \Cref{lem:same-st} is satisfied because the permutation group
is closed under composition, that is, the composition of two
permutations of $[N]$ is still a permutation of $[N]$. This property
can be violated when the test conditions on additional
events. Such an example can be found in
\citet{southworth2009properties} and is described next. Suppose that
$\bm Z$ is randomized uniformly over
$\mathcal{Z} = \{\bm z \in
  \{0,1\}^N: \bm z^T \bm 1 = N/2\}$,
so exactly half of the units are treated. Consider the so-called
``balanced permutation test'' \citep[Section 6]{efron2001empirical}
that uses the following conditioning set
\begin{equation}
  \label{eq:balanced-permutation}
  \mathcal{S}_{\bm z} = \{\bm z^{*}:
  \text{$\bm z^{*}$ is a permutation of $\bm z$ and } \bm z^T \bm z^{*}
  = N/4\}.
\end{equation}
\citet{southworth2009properties} showed that the standard theory
for permutation tests in \citet{lehmann2006testing} does not
establish the validity of the balanced permutation tests. They also provided numerical examples in which
the balanced permutation test has an inflated type I error.
Using the theory in \Cref{sec:single-crt}, we see that
\eqref{eq:balanced-permutation} clearly does not satisfy the
invariance property in \Cref{lem:same-st}. So the balanced
permutation test is not a conditional randomization test.
\cite{southworth2009properties} used the observation that
$\mathcal{S}_{\bm z}$ is not a group under balanced permutations (nor
is $\mathcal{S}_{\bm z} \cup \{\bm z\}$) to argue that the balanced
permutation test may not be valid. However, a group structure is not
necessary; in \Cref{sec:single-crt}, the only required algebraic structure
is that $\mathcal{R} = \{\mathcal{S}_{\bm z}:\bm z \in \mathcal{Z}\}$
is a partition of $\mathcal{Z}$ (or equivalently $\mathcal{R}$ is
defined by an equivalence relation). This is
clearly not satisfied by \eqref{eq:balanced-permutation}.

\subsection{Quasi-randomization tests for (conditional) independence}
\label{sec:tests-cond-indep}

Permutation tests are also frequently used to test independence of
random variables. In this problem, it is typically assumed that we observe
independent and identically distributed random variables
$(Z_1,Y_1),\dotsc,(Z_N,Y_N)$ and would like to test the null
hypothesis that $Z_1$ and $Y_1$ are independent. In the classical
treatment of this problem \citep{hajek1999theory,lehmann2006testing},
the key idea is to establish the following \emph{permutation
  principle} under the null: for all
    permutations $g:[N] \to [N]$,
\begin{equation}
  \label{eq:permutation-principle}
  (Z_1,\dotsc,Z_N,Y_1,\dotsc,Y_N) \overset{d}{=}
  (Z_{g(1)},\dotsc,Z_{g(N)},Y_1,\dotsc,Y_N).
\end{equation}
Let $\bm Z = (Z_1, \dotsc, Z_N)$, $\bm Y = (Y_1,\dotsc,Y_N)$, and
$\bm Z_{g} = (Z_{g(1)}, \dotsc, Z_{g(N)})$. Given
a test statistic $T(\bm Z, \bm Y)$, independence is rejected by the
permutation test if the following p-value is less than the
significance level $\alpha$: \qzz{(recall that there are $N!$ permutations
of $[N]$)}{}
\begin{equation}
  \label{eq:p-value-permutation}
  P(\bm Z, \bm Y) = \frac{1}{N!} \text{\qzz{$\sum_{g} $}{$\sum_{g \in \Omega_N}$}} 1_{\{T(\bm Z_{g},
    \bm Y) \leq T(\bm Z, \bm Y)\}},
\end{equation}
\qzz{}{where $\Omega_N$ be the set that collects all the $N!$
  permutations of $[N]$.}

The same notation $P(\bm Z, \bm Y)$ is used in
\eqref{eq:p-value-permutation} as it is algorithmically identical to a
conditional randomization test given the order statistics of $\bm
Z$. So it appears that
the same test can be used to solve a different, non-causal
problem---after all, no counterfactuals are involved in testing
independence. In fact, \citet{lehmann1975nonparametrics}
referred to the causal inference problem as the \emph{randomization
  model} and the independence testing problem as the \emph{population
  model}. \citet{ernst2004permutation} and
\citet{hemerik20_anoth_look_at_lady_tastin} argued that the reasoning
behind these two models is different.

However, a statistical test not only tests the null hypothesis $H_0$
but also any assumptions needed to set up the problem. For example,
the CRT described in \Cref{sec:single-crt} tests not only
the presence of treatment effect but also the assumption that
the treatment is randomized (\Cref{assump:randomization}). However,
due to physical randomization, we can treat the latter as given.

Conversely, in independence testing we may artificially define
potential outcomes as $\bm Y(\bm z) = \bm Y$
for all $\bm z \in \mathcal{Z}$, so $\bm W$ consists of many identical
copies of $\bm Y$ and the ``causal'' null
hypothesis $H_0:\bm  Y(\bm z) =\bm  Y(\bm z^{*}), \forall \bm z, \bm z^{*} \in
\mathcal{Z}$ is automatically satisfied. 
Suppose the test statistic is given by $T_{\bm z}(\bm z^{*}, \bm
W) = T(\bm z^{*}, \bm Y(\bm z^{*}))$ as in
\Cref{sec:focal-units}. Due to how the potential outcomes schedule
$\bm W$ is defined, the test statistic is simply $T(\bm z^{*}, \bm
Y)$. The CRT then tests $\bm Z \independent \bm
W$ in \Cref{assump:randomization}, which is equivalent to $\bm Z
\independent \bm Y$. When $\mathcal{S}_{\bm Z}$ is given by
all the permutations of $\bm Z$ as in \eqref{eq:sz-permutation}, $\bm
Z^{*} \mid \bm Z^{*} \in \mathcal{S}_{\bm Z}$ has the same
distribution as $\bm Z_{g}$ where $g$ is a random
permutation. These observations show that the permutation
test of independence is identical to the permutation test for the artificial ``causal'' null hypothesis.

So permutation tests of  treatment effect and independence
are two ``sides'' of the same ``coin''. They are algorithmically the
same, but differ in what is regarded as the presumption
and what is regarded as the hypothesis being tested. Rather than
distinguishing them according to the type of ``model'' (randomization
or population), we believe that the more fundamental difference is the
nature of randomness used in each test. In testing
treatment effect, inference is entirely based on the randomness
introduced by the experimenter and is thus a randomization test. In
testing independence, inference is based on the
permutation principle \eqref{eq:permutation-principle} that follows
from a theoretical model, so the same permutation test is a
quasi-randomization test in our terminology.




Recently, there is a growing interest in using quasi-randomization tests
for conditional independence
\citep{berrett2020conditional,candes2018panning,katsevich2020theoretical,liu2020fast}. Typically,
it is assumed that we have independent and identically distributed observations
$(Z_1,Y_1,X_1),\dotsc,(Z_n,Y_n,X_n)$ and would like to
test $Z_1 \independent Y_1 \mid X_1$. This can be easily incorporated
in our framework by treating $\bm X = (X_1,\dotsc, X_n)$ as fixed; see
the last paragraph in \Cref{sec:potent-outc-fram}. In this case, the
quasi-randomization distribution of $\bm Z = (Z_1,\dotsc,Z_n)$ is given by
the conditional distribution of $\bm Z$ given $\bm X$, and it is
straightforward to construct a quasi-randomization p-value
\citep[see e.g.][Section
4.1]{candes2018panning}. \citet{berrett2020conditional} extended this
test by further conditioning on the order statistics of $\bm Z$,
resulting in a permutation test.

As a remark on the terminology, the test in the last paragraph was
referred to as the ``conditional randomization test'' by
\cite{candes2018panning} because the test
is conditional on $\bm X$. However, such conditioning is fundamentally
different from post-experimental conditioning (such as conditioning on
$\mathcal{S}_{\bm Z}$), which is used in \Cref{sec:single-crt} to
distinguish conditional from unconditional randomization tests. When
$\bm Z$ is randomized according to $\bm X$,
conditioning on $\bm X$ is mandatory in randomization inference
because it needs to use the randomness introduced by the
experimenter. On the other hand,
further conditioning on $\mathcal{S}_{\bm Z}$ or more generally $G =
g(\bm Z, V)$ in \Cref{sec:randomized-crts} is
introduced by the analyst to improve the power or make the p-value
computable. For this reason, we think it is best to refer to the test in
\citet{candes2018panning} as a unconditional quasi-randomization test and
the permutation test in \citet{berrett2020conditional} as a conditional
quasi-randomization test.

\subsection{Covariate imbalance and adaptive randomization}
\label{sec:form-part-covar}

\citet{morgan2012rerandomization} proposed to rerandomize the
treatment assignment if some baseline covariates are not well
balanced. \qzz{}{\citet{li19_reran_regres_adjus} showed that the asymptotic
distribution of standard regression-adjusted estimators in this design
is a mixture of a normal distribution and a truncated normal
distribution, whose variance is always no larger than that in the
standard completely randomzied design.}

Notice that the meaning of ``rerandomization'' here is completely
different from that in ``rerandomization test''
\citep{brillinger1978role,gabriel83_reran_infer_regres_shift_effec},
which \qzz{is a misnomer for}{emphasizes on the Monte-Carlo
  approximation} to a randomization test. The key insight of
\cite{morgan2012rerandomization} is
that the experiment should then be analyzed with the rerandomization
taken into account. More
specifically, rerandomization is simply a rejection sampling algorithm
for randomly choosing $\bm Z$ from the subset
$ \mathcal{Z} = \{\bm z: g(\bm z) \leq \eta\},$
where $g(\bm z)$ measures the covariate imbalance implied by the
treatment assignment $\bm z$ and $\eta$ is the experimenter's
tolerance of covariate imbalance. Therefore, we simply need to use the randomization
distribution over $\mathcal{Z}$ to carry out the randomization
test. The same reasoning applies to other adaptive trial designs, see
\cite{rosenberger18_random} and references therein.

Another way to deal with unlucky draws of treatment assignment is to
condition on the covariate imbalance $g(\bm Z)$
\citep{zheng08_multi_center_clinic_trial,hennessy2016conditional}. This inspired
\Cref{prop:condition-on-function} in \Cref{sec:conditioning-sets}. In
our terminology, this is a conditional randomization test because the
analyst has the liberty to choose which function $g(\bm Z)$ to
condition on. On the other hand, the randomization test proposed by
\citet{morgan2012rerandomization} is unconditional.



\subsection{Conformal prediction}
\label{sec:conformal-prediction}

Conformal prediction is another topic related to randomization
inference that is receiving rapidly growing interest in statistics and
machine learning
\citep{vovk2005algorithmic,shafer2008tutorial,lei13distribution,lei18distribution}. Consider
a typical regression problem where the data points
$(X_1,Y_1),\dotsc,(X_{N},Y_{N})$ are drawn exchangeably from the same
distribution and $Y_N$ is unobserved. We would like to construct a prediction interval
$\hat{\mathcal{C}}(X_N)$ for the next observation $Y_N$ such
that $ \mathbb{P}(Y_N \in \hat{\mathcal{C}}(X_N)) \leq 1 - \alpha.$
The key idea of conformal prediction is that the exchangeability of
$(X_1,Y_1),\dotsc, (X_N,Y_N)$ allows us to use permutation tests for
the null hypothesis $H_0:Y_N = y$ for any fixed $y$. More
concretely, we may fit \emph{any} regression model to
$(X_1,Y_1),\dotsc,(X_{N-1},Y_{N-1}),(X_N,y)$ and let the p-value be
one minus the percentile of the absolute residual of $(X_N,y)$ among
all $N$ absolute regression residuals. A large
residual of $(X_N,y)$ indicates that the prediction $(X_N,y)$
``conforms'' poorly with the other observations, so $H_0:Y_N = y$ should be
rejected.

Conformal prediction can be investigated in the framework in
\Cref{sec:single-crt} by viewing sampling as a
form of quasi-randomization. Implicitly, conformal prediction
conditions on the unordered data points
$(X_1,Y_1),\dotsc,(X_N,Y_N)$. Let the ``treatment variable'' $Z$ be the
order of the observations, which is a random permutation of $[N]$ or
equivalently a random bijection from $[N]$ to $[N]$. The
``potential outcomes'' are then given by
\begin{equation}
  \label{eq:conformal-counterfactual}
  \bm Y(z)  = \left( (X_{z(1)},Y_{z(1)}),\dotsc,(X_{z(N)},Y_{z(N)}) \right),
\end{equation}
The notation $\bm Y$ is overloaded here to be consistent with
\Cref{sec:single-crt}. The ``potential outcomes schedule'' in our setup is
$\bm W = \left\{\bm Y(z) : \text{ $z$ is a permutation of $[N]$}\right\}$.

The null hypothesis in conformal prediction is $H_0:Y_{N} = y$. This
is a sharp null hypothesis, and our theory in
\Cref{sec:single-crt} can be used to construct a
quasi-randomization test. Notice that randomization in our theory ($Z
\independent \bm W$ in \Cref{assump:randomization}) is
justified here by the theoretical assumption that
$(X_1,Y_1),\dotsc,(X_N,Y_N)$ are exchangeable. In other words, we may
view random sampling as randomizing the order of the observations. But
unless physical randomization is involved in sampling, this is
only a quasi-randomization test.

The above argument can be extended to allow ``covariate shift'' in the
next observation $X_N$, in the sense that $X_N$ is drawn from a
distribution different from $X_1,\dotsc,X_{N-1}$
\citep{tibshirani2019conformal,hu20distribution}. Again, the key idea
is to consider the quasi-randomization formulation of sampling. For
this we need to consider a (potentially infinite) super-population $(X_i,Y_i)_{i
  \in \mathcal{I}}$. The ``treatment'' $Z: [N] \to \mathcal{I}$
selects which units are observed. For example, $Z(1) = i$ means that
the first data point is $(X_i,Y_i)$. The ``potential outcomes'' are still
given by \eqref{eq:conformal-counterfactual}. By conditioning on
the image of $Z$, we can obtain a conditional quasi-randomization test
for the unobserved $Y_{z(N)}$. Covariate shift in $X_N$ can be easily
incorporated by conditioning on $(X_{Z(1)},\dotsc, X_{Z(N)})$ and deriving the
conditional distribution of $Z$. For
example, suppose 
\begin{equation}
  \label{eq:cov-shift-dist}
  \begin{split}
  &\mathbb{P}(Z(k) = i) \propto \pi_1(X_i)~\text{for}~k=1,\dotsc,N-1,i \in
  \mathcal{I}, \\
  &\mathbb{P}(Z(N) = i) \propto \pi_2(X_i)~\text{for}~i \in
  \mathcal{I}.
  \end{split}
\end{equation}
The image $\{Z(1),\dotsc,Z(N)\}$ is a multiset
$\mathcal{U} = \{u_1,\dotsc,u_N\}$ of $N$ possibly duplicated
units. Let $\bm X = (X_{u_1},\dotsc,X_{u_N})$ and $z:[N] \to
\mathcal{U}$ be any permutation of $\mathcal{U}$. The
conditional quasi-randomisation distribution of $Z$ is then given by
\begin{align} \label{eq:cov-shift-random-dist}
  \P\left(Z = z \mid \bm X, \mathcal{U}\right)
    \propto 
      \pi_2(X_{z(N)})  \prod_{j =1}^{N-1}\pi_1(X_{z(j)})         
   \propto \frac{\pi_2(X_{z(N)})}{\pi_1(X_{z(N)})},~i=1,\dotsc,N,
\end{align}
by using the fact that the product $\prod_{j=1}^{N}\pi_1(X_{z(j)}) =
\prod_{j=1}^N$ $\pi_1(X_{u_j})$ is a function of $\mathcal{U}$.
By normalizing the right hand side, we arrive at
\[
  \P\left(Z = z \mid \bm X,\mathcal{U}\right)
  =
  \frac{\pi_2(X_{z(N)})/\pi_1(X_{z(N)})}{\sum_{j=1}^N\pi_2(X_{u_j})/\pi_1(X_{u_j})},~ i=1,\dotsc,N.
\]
The last display equation was termed ``weighted exchangeability'' in
\cite{tibshirani2019conformal} and is particularly convenient because
the conditional quasi-randomization distribution only depends on
the density ratio. However, we would like to stress again
that exchangeability (unweighted or weighted) is not essential in
constructing a randomization test. All that is required is the
distribution of the ``treatment'' $\bm Z$. For example, sampling
schemes that are more general than \eqref{eq:cov-shift-dist} can be
allowed in principle, although the conditional quasi-randomization
distribution might not be as simple as \eqref{eq:cov-shift-random-dist}.


%% file: 6_Discussion_Revised.tex
To reiterate the main thesis of this article, a randomization test
\qzz{is}{should be}---as its name suggests---a hypothesis test based on
the physical act of randomization. In view of our trichotomy of the
randomness in data and statistical tests, this point is made precise
by conditioning on the potential outcomes schedule that represents the
randomness in the nature. In this framework, the
analysts are allowed to choose a test statistic, condition on certain
aspects of the treatment, and even introduce further
randomization. \qzz{}{Contrary to common beliefs, randomization tests
  are not the same as, or a superset/subset of, permutation
  tests. They speak to fundamentally different aspects of a
  test---what randomness the test is based on and what algorithm can
  be used to compute the test.}

Without physical randomization, we believe it is best to refer to the
same test as a quasi-randomization test, as the statistical inference
will be inevitably based on some unverifiable modeling
assumptions.\footnote{Philosophically, it may be argued that true
  physical (``ontological'') randomization does not exist; perhaps
  which ball is drawn from an urn is not random to a
  superhuman. Still, it is much better to base inference on the randomness (we
  think) we introduce and understand well.} After all, it is easy for
us statisticians to assume the data
are independent and identically distributed and forget that this is
only a theoretical model that requires justification in practice. On
this, \citet{fisher1922mathematical} offered a sobering remark:
``The postulate of randomness thus resolves itself into the question,
`Of what population is this a random sample?' which must frequently be
asked by every practical statistician.''

In \Cref{sec:single-crt} we have described three
perspectives on conditioning in randomization tests: conditioning on a
set of treatment assignments, conditioning on a $\sigma$-algebra, and
conditioning on a random variable. They are useful
for different purposes: the first perspective is useful when the null
hypothesis is partially sharp and one needs to construct a computable
p-value; the second perspective is useful to describe the
conditioning structure of multiple CRTs; and the third perspective is the most
interpretable and allows us to consider post-experimental
randomization. We have collected some useful methods to construct CRTs in
\Cref{sec:practical-methods}. In most of our development there,
conditioning is presented as a technique to obtain computable
randomization tests. But conditioning can
improve power \citep{hennessy2016conditional} and lead to good
frequentist properties \citep{lehmann2006testing}. Theoretical guidance
on designing the conditioning event is an interesting direction
for future work.

So which statistical test should be applied in practice---a
randomization test, a quasi-randomisation test, or a model-based test
(such as the $F$-test in linear models)? First of all, a randomization
test is not always possible unless there is physical act of
randomization; when the randomization test is feasible, it provides a
pristine way of inference for a very specific task---testing a sharp
null hypothesis. Thus in a randomized experiment, there seems to be no
reason to not report the result of a randomization test \cite[for more
discussion,
see][]{rosenberger18_random,rubin80comment}. Quasi-randomization
and model-based tests may have better power, but they
are sensitive to model misspecification or
violations of any assumption involved in justifying the
quasi-randomisation (such as exchangeability), thus careful arguments
are needed to justify their utility. When no randomization test is
available (e.g. in construction of prediction intervals without
experimental data), the choice between quasi-randomisation and
model-based tests will rest on how reasonable the theoretical
assumptions are in a practical problem and how easily these tests can
be computed.


%% file: appendix.tex
\section{Proof of \Cref{lemma:computable}}

\begin{proof}
  By \Cref{def:hypothesis}, $\bm W_{\bm Z}=(\bm Y_{\mathcal{H}(\bm Z, \bm
    z^{*})}(\bm z^{*}): \bm z^{*} \in \mathcal{S}_{\bm
    Z})$ denote all the potential outcomes  imputable from $\bm Y = \bm Y(\bm Z)$
  under $H$. Then $\bm W_{\bm Z}$ is fully determined by $\bm Z$ and
  $\bm Y$. By assumption,  $T_{\bm Z}(\bm z^{*}, \bm W)$ only depends
  on $\bm W$ through $\bm W_{\bm Z}$ for
  all $\bm z^*\in \mathcal{S}_{\bm Z}$
  . Thus, $T_{\bm Z}(\bm z^{*}, \bm W)$ is a function of $\bm
  z^{*}$, $\bm Z$, and $\bm Y$. With an abuse of notation we denote it
  as $T_{\bm Z}(\bm z^{*}, \bm Y)$. By the definition of the p-value
  in \eqref{eq:p-value} and $\bm Z^{*} \independent \bm Z \independent
  \bm W$,
  \begin{align*}
    P(\bm Z, \bm W) &= \mathbb{P}^{*} \{T_{\bm Z}(\bm Z^{*}, \bm W)
                      \leq T_{\bm Z}(\bm Z, \bm W) \mid \bm Z^{*}\in \mathcal{S}_{\bm Z},
                      \bm W \} \\
                    &= \mathbb{P}^{*} \{T_{\bm Z}(\bm Z^{*}, \bm Y)
                      \leq T_{\bm Z}(\bm Z, \bm Y) \mid \bm Z^{*}\in \mathcal{S}_{\bm Z},
                      \bm W \} \\
                    &= \mathbb{P}^{*} \{T_{\bm Z}(\bm Z^{*}, \bm Y)
                      \leq T_{\bm Z}(\bm Z, \bm Y) \mid \bm Z^{*}\in \mathcal{S}_{\bm Z}
                      \}
  \end{align*}
  is a function of $\bm Z$ and $\bm Y$ (since $\P^{*}$ only depends on
  the known density $\pi(\cdot)$).
\end{proof}

\section{Proof of \Cref{thm:valid}}

\begin{proof}
  We first write the p-value \eqref{eq:p-value} as a probability
  integral transform. For any fixed $\bm z \in \mathcal{Z}$, let
  $F_{\bm z}(\cdot;\bm W)$ denote the distribution function of $T_{\bm
    z}(\bm Z, \bm W)$ given $\bm W$ and $\bm Z \in \mathcal{S}_{\bm
    z}$. Given $\bm Z \in \mathcal{S}_{\bm
    z}$ (so $\mathcal{S}_{\bm Z} = \mathcal{S}_{\bm z}$ and $T_{\bm Z}
  = T_{\bm z}$ by \Cref{lem:same-st}), the p-value can be written
  as
  \[
    P(\bm Z, \bm W) = F_{\bm z}(T_{\bm z}(\bm Z, \bm W);\bm W).
  \]
  To prove \eqref{eq:stoc-domi}, we can simply use the following
  result in probability theory: let $T$ be a real-valued random
  variable and $F(t) = \P(T \le t)$ be its distribution function, then
  $\mathbb{P}(F(T) \leq \alpha) \leq \alpha$ for all $0 \leq \alpha
  \leq 1$. 

  If the p-value is computable, we have $P(\bm Z, \bm W) = P(\bm Z, \bm
  Y)$ by \Cref{lemma:computable}. By the law of total probability, for
  any $\alpha \in [0,1]$,
  \begin{align*}
    \mathbb{P}\left\{ P(\bm Z, \bm Y)\leq \alpha \mid \bm W \right\} &=
                                                                       \sum_{m=1}^M \mathbb{P}\left\{ P(\bm Z, \bm Y)\leq \alpha \mid \bm
                                                                       Z \in \mathcal{S}_m, \bm W \right\} \mathbb{P}(\bm Z \in
                                                                       \mathcal{S}_m \mid \bm W) \\
                                                                     &\leq
                                                                       \sum_{m=1}^M
                                                                       \alpha
                                                                       \,\mathbb{P}(\bm
                                                                       Z \in \mathcal{S}_m \mid
                                                                       \bm
                                                                       W)
                                                                       =
                                                                       \alpha.
  \end{align*}
\end{proof}

\section{Proof of \Cref{cor:valid-2}}

\begin{proof}
  This follows from the observation that once $G$ is given, this is
  simply a unconditional randomization test. The p-value is computed using
  $\bm Z^{*} \sim \pi(\cdot \mid G)$, so given $G$, we can write
  $P(\bm Z, \bm W; G)$ as a probability integral transform as in the
  proof of \Cref{thm:valid}.
\end{proof}
